# A High Precision Time Measurement Method Based on Frequency-domain Phase-Fitting for Nuclear Pulse Detection

Jianjun Wang, Zhaohui Bu, Zhao Wang, Jincheng Xu, Liguo Zhou, Qibin Zheng

*Abstract*—This paper proposes a high-precision time measurement method based on digital frequency-domain phase-fitting (DFPF) by using the digitized nuclear pulses. The averaging effect inherent in the frequency-domain cross-correlation and phase-fitting processes effectively minimizes measurement errors, thereby ensuring high precision and resolution in time interval measurements. In this paper, the theory of this DFPF-based time measurement method is analyzed, and an electronics prototype is designed to validate the feasibility of the proposed method by utilizing ADCs for pulse digitization and an FPGA for phase fitting implementation. The test results indicate that, under ideal conditions with a signal-to-noise ratio (SNR) of 64 dB, this method achieves time measurement precisions of 50 ps, 18 ps, and 2.9 ps RMS, corresponding to different Gaussian pulse widths and sampling rates of 118 ns at 40 MSPS, 10 ns at 100 MSPS, and 3 ns at 500 MSPS, respectively. The precision improves with increasing pulse bandwidth. Furthermore, in practical cosmic ray tests, the method achieved favorable timing performance with a precision of 1.7 ns RMS. These results demonstrate that this proposed method has the potential to be a high-precision time measurement for particle detection and is equally applicable to other advanced time measurement scenarios.

*Index Terms*—time measurement, frequency-domain phase-fitting, nuclear pulse detection

## I. INTRODUCTION

High-precision time measurement is critical in nuclear physics and its related applications, with significant impacts across various fields such as particle physics experiments [1], medical applications [2], and radiation monitoring [3]. Accurate timing is essential for characterizing nuclear reactions, improving time-of-flight (TOF) measurements, enhancing energy resolution, and optimizing instrument performance. Many existing nuclear detection systems utilize conventional approaches, primarily integrating time-to-digital converters (TDCs) with analog-to-digital converters (ADCs) in circuit board designs [4-7]or chip-level integrations[8, 9]. In these systems, time measurement and energy measurement are implemented as relatively independent units. However, as the scale and precision requirements of fundamental experiments [10], medical instruments [11] and other applications continue to grow, there is an increasing demand for simpler and more integrated solutions. Digital timing methods, which extract the arrival time of a nuclear pulse from its digitized waveform, have garnered significant attention [12-16], offering several advantages over conventional approaches. Digital timing enables simultaneous pulse waveform discrimination and energy measurement [17], reduces drift [18], minimizes the need for preamplifier and shaping electronics [19], and enhances cost efficiency in large-scale systems [12, 20].These advantages highlight digital timing as a promising solution for nuclear pulse detection, yet a key limitation is its finite time precision.

Recent advancements in high-speed ADCs and digital processing techniques, such as digital constant fraction discrimination (DCFD), along with various interpolation methods, waveform fitting, and noise-optimized timing filters, have significantly enhanced timing performance and robustness. However, challenges such as electronic noise, achieving high time precision with low-speed ADCs, and computational complexity still persist and must be further addressed to meet the growing demands for widespread applications. Interestingly, the current state of research indicates that ADC-based time measurement in the field of nuclear detection is primarily conducted in the time domain. Moreover, it is well-established that, as long as the Nyquist sampling theorem is satisfied, all spectral information of the pulse signal, including its time information, is captured. Therefore, exploring novel, simple, and efficient high-precision time measurement methods in the frequency domain, which only require low-sampling-rate ADCs for waveform digitization, presents a promising avenue for future research.

This paper explores a high-precision time measurement method based on digital frequency-domain phase-fitting (DFPF) for nuclear pulse detection. In this method, the time information of digitized pulses is estimated by performing phase-fitting analysis on their frequency-domain cross-correlation function.

Manuscript received xxx xx, 2024; revised xxx xx, xxx2024. This project is supported in part by grants from National Natural Science Foundation of China (No. 12105177 and 11904423) and a grant from National Key R&D Program of China (No. 2023YFF0719200), and support also from Shanghai Key Laboratory of Particle Physics and Cosmology (No. 22DZ2229013-1). (Corresponding author: Qibin Zheng;)

Jianjun Wang, Zhaohui Bu, Jincheng Xu, Liguo Zhou and Qibin Zheng are with the Quantum Medical Sensing Laboratory and School of Health Science and Engineering, University of Shanghai for Science and Technology, Shanghai 200093, China (e-mail: qbzheng@usst.edu.cn).

Zhao Wang is with the Southern University of Science and Technology, Shenzhen 518055, China, and International Quantum Academy, Shenzhen 518048, China.

Jianjun Wang and Zhaohui Bu contributed equally to this work.

The averaging effect inherent in the cross-correlation and phase-fitting processes effectively minimizes measurement errors, ensuring high precision and resolution in time interval measurements. To validate the feasibility of the proposed method, an electronic prototype is designed, which utilizes an ADC to digitize the pulses, and the time measurement algorithm based on phase fitting is implemented within an FPGA. A high-precision time measurement can be achieved without requiring an independent time circuit. Moreover, compared to waveform fitting, the DFPF-based high-precision time measurement method only needs to satisfy the Nyquist sampling theorem, effectively preventing data redundancy. Building upon this, we have conducted cosmic ray testing cosmic ray testing to assess the applicability of the method. The test results indicate that this method can enable the development of high-precision readout electronics for the application in nuclear pulse detection and other advanced time measurement systems.

This paper is organized as follows: Section II introduces the principles of DFPF-based time measurement. Additionally, the method is subjected to measurement error analysis. In Section III, simulation tests are developed to validate this method. In Section IV, a prototype is designed and implemented, followed by an electronics test and cosmic ray test to assess its performance. A summary of this research and a discussion of future improvements are given in Section V.

## II. THEORETICAL ANALYSIS

### A. Time Measurement Method Based on DFPF

In a typical nuclear detection system, nuclear pulse signals are usually amplified and shaped by the front-end amplifier (FEA) after passing through the detector, resulting in a series of pulse signals with similar spectral information. After FEA processing, the procedure for the time measurement method is depicted in Fig. 1. Two nuclear pulse signals with a time interval $\tau$ are digitized by the ADCs and subsequently sent to the processor for further analysis. Initially, the digitized signals are transformed into frequency domain signals through discrete Fourier transform (DFT). Then, the cross-correlation operations are applied to the signals to extract the frequency and phase information of their cross-correlation functions. Theoretical analysis indicates that a linear relationship exists between the phase and frequency, with the slope of this relationship being dependent on time intervals. Consequently, the time measurement is achieved through the application of a phase-fitting algorithm.

The nuclear pulse signals $s_1(t)$, $s_2(t)$ in Fig. 1 with a time interval $\tau$ can be expressed as:

$$s_1(t) = s(t) \quad (1)$$
$$s_2(t) = s(t-\tau). \quad (2)$$

The digitized signal $x_1(nT_s)$, $x_2(nT_s)$ obtained after sampling with the ADCs can be expressed as:

$$x_1(nT_S) = s_1(nT_S) + w_1(nT_S) \quad (3)$$
$$x_2(nT_S) = s_2(nT_S) + w_2(nT_S) \quad (4)$$

where $w_1(nT_s)$, $w_2(nT_s)$ represent the noise signal introduced by the circuit. Furthermore, according to the correlation theorem [21], performing cross-correlation on the obtained digital signals and applying the DFT yields:

$$R(k) = X_1(k)X_2^*(k)$$
$$= S_1(k)S_2^*(k) + S_1(k)W_2^*(k) + S_2^*(k)W_1(k) + W_1(k)W_2^*(k). \quad (5)$$

In the expression, $X_1(k)$, $X_2(k)$, $S_1(k)$, $S_2(k)$, $W_1(k)$, $W_2(k)$ represent the spectra of $x_1(nT_s)$, $x_2(nT_s)$, $s_1(nT_s)$, $s_2(nT_s)$, $w_1(nT_s)$, $w_2(nT_s)$, here $k$ is the frequency.

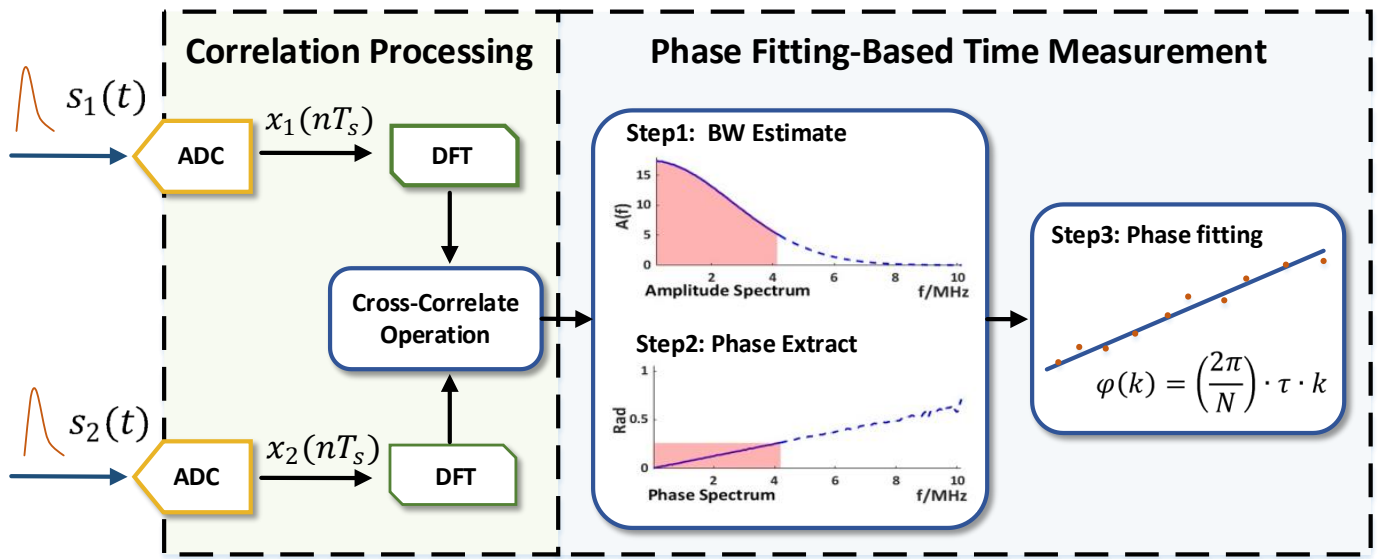

Fig. 1. Block diagram of high precision time measurement of nuclear pulse events based on the DFPF method. When nuclear pulses strike the detector, they are shaped into quasi-Gaussian pulses s₁(t), s₂(t) through analog frontend processing. x₁(nT_S) and x₂(nT_S) represent the digital signals of the pulses, and T_S is the sampling period of the ADCs. $\varphi(k)$ is the phase angle of in the bandwidth and k is the frequency.

The noise is mutually independent and unrelated to the pulse signal when the number of sampling points N is sufficiently large. Then the cross-correlation can be approximately written as:

$$R(k) \approx S_1(k)S_2^*(k) = |S(k)|^2 \exp(j\frac{2\pi}{N}k\tau). \quad (6)$$

The phase of $R(k)$ can be expressed as:

$$\varphi(k) = \frac{2\pi}{N}k\tau. \quad (7)$$

It means the phase $\varphi(k)$ is proportional to the frequency $k$ and the delay time $\tau$. If we perform the linear regression method, the delay time $\tau$ can be estimated as the slope of the linear relationship between the phase $\varphi(k)$ and the frequency $k$, denoted as $\hat{\tau}$.

*B. Error Analysis*

Furthermore, the error of estimation value $\hat{\tau}$ is analyzed. The relationship between the phase $\varphi_n$ and frequency points $f_n$ can be represented as:

$$\varphi_n = 2\pi f_n \tau + \phi_0 \ (n=0,1,\cdots,N-1) \quad (8)$$

where $\phi_0$ is the initial phase, and $N$ refers to the number of sampling points.

Let $\vec{y}$ denote the observation vector, $\vec{x}$ denote the vector of estimated parameters, and $J$ indicate the corresponding Jacobian determinant. Therefore, the following relationship can be established:

$$\vec{y} = J\vec{x} \quad (9)$$

where

$$\vec{y} = \begin{bmatrix} \varphi_0 \\ \varphi_1 \\ \vdots \\ \varphi_{N-1} \end{bmatrix}, \quad J = \begin{bmatrix} 2\pi f_0 & 1 \\ 2\pi f_1 & 1 \\ \vdots & \vdots \\ 2\pi f_n & 1 \end{bmatrix}, \quad \vec{x} = \begin{bmatrix} \tau \\ \phi_0 \end{bmatrix}. \quad (10)$$

Let $\sigma_{\varphi_n}$ represent the phase error of $\varphi_n$, with the error matrix $V$ given by:

$$V = \begin{bmatrix} \sigma_{\varphi_0}^2 & & & 0 \\ & \sigma_{\varphi_1}^2 & & \\ & & \ddots & \\ 0 & & & \sigma_{\varphi_{N-1}}^2 \end{bmatrix}. \quad (11)$$

The mean square error matrix *MSE* for the least-squares estimates $\hat{\tau}$ and $\hat{\phi}_0$ can be expressed as:

$$MSE = \left[J^T \ V^{-1} \ J\right]^{-1}$$

$$= \frac{1}{\sum_{n=0}^{N-1}\frac{(2\pi f_n)^2}{\sigma_{\varphi_n}^2} \cdot \sum_{n=0}^{N-1}\frac{1}{\sigma_{\varphi_n}^2} - (\sum_{n=0}^{N-1}\frac{2\pi f_n}{\sigma_{\varphi_n}^2})^2} \cdot \begin{bmatrix} \sum_{n=0}^{N-1}\frac{1}{\sigma_{\varphi_n}^2} & -\sum_{n=0}^{N-1}\frac{2\pi f_n}{\sigma_{\varphi_n}^2} \\ -\sum_{n=0}^{N-1}\frac{2\pi f_n}{\sigma_{\varphi_n}^2} & \sum_{n=0}^{N-1}\frac{(2\pi f_n)^2}{\sigma_{\varphi_n}^2} \end{bmatrix} \quad (12)$$

In (12), as described by Takahashi et al. (2000), $\sigma_{\varphi_n}^2$ can be approximated as a constant value $\sigma_\varphi^2$ which is inversely proportional to the signal-to-noise ratio at the corresponding frequency point of the cross-correlation function ($SNR_n$) [22]:

$$\sigma_\varphi^2 = \frac{A}{SNR_n} \quad (13)$$

where $A$ is the coefficient. The $SNR_n$ is found by dividing the signal-to-noise ratio ($SNR$) of the output signal from the cross-correlation function ($SNR_o$) into N parts [22]:

$$SNR_o = N \cdot SNR_n. \quad (14)$$

Then the *MSE* can be represented as:

$$MSE = \frac{\sigma_\varphi^2}{4\pi^2 N^2 \left\{\frac{1}{N}\sum_{n=0}^{N-1}f_n^2 - \left(\frac{1}{N}\sum_{n=0}^{N-1}f_n\right)^2\right\}} \cdot \begin{bmatrix} N & -2\pi\sum_{n=0}^{N-1}f_n \\ -2\pi\sum_{n=0}^{N-1}f_n & 4\pi^2\sum_{n=0}^{N-1}f_n^2 \end{bmatrix}$$

$$\approx \frac{A}{4\pi^2 N \cdot SNR_o \cdot \left\{\frac{1}{N}\sum_{n=0}^{N-1}f_n^2 - \left(\frac{1}{N}\sum_{n=0}^{N-1}f_n\right)^2\right\}} \cdot \begin{bmatrix} N & -2\pi\sum_{n=0}^{N-1}f_n \\ -2\pi\sum_{n=0}^{N-1}f_n & 4\pi^2\sum_{n=0}^{N-1}f_n^2 \end{bmatrix}$$

(15)

where $\frac{1}{N}\sum_{n=0}^{N-1}f_n$ can be represented as the sample mean of the frequency $\bar{f}$. The sample variance of the frequency $\sigma_f^2$ is given by:

$$\sigma_f^2 = \frac{1}{N}\sum_{n=0}^{N-1}\left(f_n - \bar{f}\right)^2$$
$$= \frac{1}{N}\sum_{n=0}^{N-1}f_n^2 - \left(\frac{1}{N}\sum_{n=0}^{N-1}f_n\right)^2. \quad (16)$$

Substituting (16) into (15), the *MSE* can be expressed as:

$$MSE = \frac{A}{4\pi^2 SNR_o \cdot \sigma_f^2} \bullet \begin{bmatrix} 1 & -2\pi\bar{f} \\ -2\pi\bar{f} & 4\pi^2\bar{f}^2 \end{bmatrix}. \quad (17)$$

The error $\varepsilon$ associated with the estimate $\hat{\tau}$ is expressed as:

$$\varepsilon = \hat{\tau} - \tau = \left(\frac{A}{4\pi^2 \cdot SNR_o \cdot \sigma_f^2}\right)^{\frac{1}{2}}. \quad (18)$$

Furthermore, for bandpass signals, $\sigma_f^2$ can be approximated as:

$$\sigma_f^2 \approx \frac{1}{f_B}\int_0^{f_B}\left(f - \frac{f_B}{2}\right)^2 df = \frac{f_B^2}{12} \quad (19)$$

where $f_B$ represents the signal bandwidth. Substituting (19) into (18) yields:

$$\varepsilon \approx \frac{\sqrt{3 \cdot A}}{\pi \cdot f_B \cdot \sqrt{SNR_o}}. \quad (20)$$

The detector combined with the preamplifier circuit can be approximated as a linear time-invariant (LTI) system under normal operating conditions [23]. When specific nuclear particles are received, the output is the result of the convolution integral between the input signal and the system's impulse response [24]. In this case, both the time-domain waveform and the bandwidth $f_B$ of the output signal are determined. It can be inferred from (20) that $\varepsilon$ decreases as the $SNR_o$ and $f_B$ increases, thereby leading to improved measurement precision. In general, as the SNR of the input signal increases, the $SNR_o$ also improves accordingly [25].

## III. SIMULATION

In order to validate this method, a simulation model was developed on the MATLAB platform. As shown in Fig. 2, the Gaussian pulse signals with noise and the ideal Gaussian pulse signals were used as test signals for theoretical calculations and simulation verification. By subtracting the cross-correlation output signal power $P_1$ of the ideal signals from the cross-correlation output signal power $P_2$ of the signals with noise, the noise power is obtained. Subsequently, the $SNR_o$ is calculated under the noise conditions. and this value was substituted into (20) to determine the theoretical precision. Additionally, the standard deviation (STD) of the estimated time results is calculated through Monte Carlo simulations.

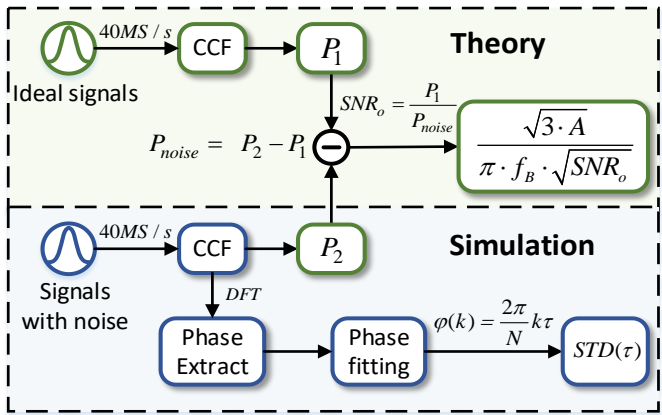

Fig. 2. The process of the theoretical calculation and simulation. $P_1$ refers to the output power of the cross-correlation function (CCF) for the ideal pulse signals. $P_2$ refers to the output power of the CCF for signals with noise. The noise power $P_{noise}$ is obtained by subtracting $P_1$ from $P_2$. Then, the $SNR_o$ can be calculated and used to determine theoretical precision.

In this simulation, the nuclear pulse was modeled as a Gaussian pulse signal with certain parameters. Firstly, a Gaussian pulse with a standard deviation ($\sigma$) of 50 ns, corresponding to a full width at half maximum (FWHM) of 117.750 ns, and an SNR of 100 dB was selected as the ideal signal. These pulses are sampled at a rate of 40 MSPS, a set of 512 samples was collected in one simulating turn. By comparing the signal power of the cross-correlation function of these input pulses with that of ideal Gaussian pulses, the $SNR_o$ was calculated to be 111.35 dB. The theoretical STD using (20) was $0.196\sqrt{A}$ ps RMS. Additionally, STD was found to be 0.52 ps RMS following 10,000 statistical results from Monte Carlo simulations under the same input conditions. Therefore, coefficient A was different from various experiments, without the loss of the generality. The value A can be approximated as 7.00 in this simulation.

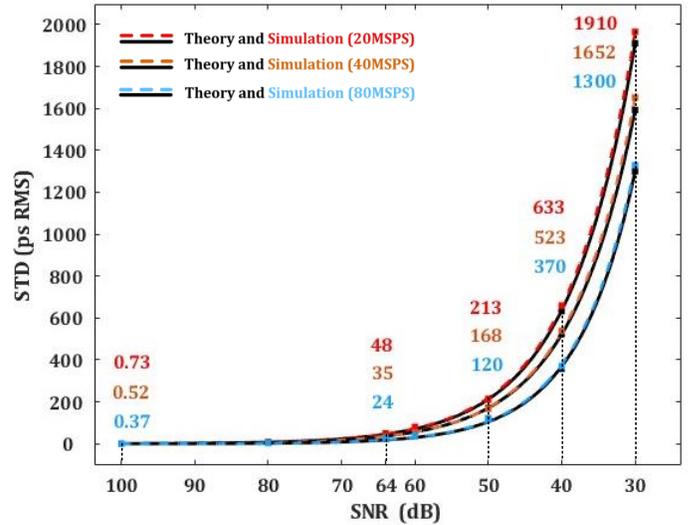

Fig. 3. The trend of the theoretical precision and the Monte Carlo simulation precision under a pulse with of 117.750 ns, an SNR ranging from 30 dB to 100 dB, and with sampling rates of 20 MSPS, 40 MSPS, and 80 MSPS.

TABLE I
DFPF-BASED TIME MEASUREMENT PRECISION UNDER DIFFERENT SIMULATION CONDITIONS

| Pulse Width (ns) | SNR (dB) | Precision (ps rms) | | |
|---|---|---|---|---|
| | | ADC0 | ADC1 | ADC2 |
| | | 80 MSPS | 40 MSPS | 20 MSPS |
| 117.750 | 64 | 24 | 35 | 48 |
| | 50 | 120 | 168 | 213 |
| | 40 | 370 | 523 | 633 |
| | 30 | 1300 | 1652 | 1910 |
| | | 500 MSPS | 250 MSPS | 100 MSPS |
| 10.127 | 64 | 5.5 | 7.8 | 13 |
| | 50 | 29 | 47 | 52 |
| | 40 | 98 | 146 | 163 |
| | 30 | 312 | 456 | 523 |
| | | 1500 MSPS | 1000 MSPS | 500 MSPS |
| 2.826 | 64 | 1.1 | 1.4 | 1.8 |
| | 50 | 6.3 | 7.6 | 9.4 |
| | 40 | 19 | 23 | 31 |
| | 30 | 65 | 84 | 100 |

Then, we introduced more Gaussian white noise into the input signals to further analyze the impact of *SNR* on time precision. Theoretical calculations and simulation were performed for input signals with SNR conditions ranging from 30 dB to 100 dB. Meanwhile, the consistency between Monte Carlo simulation results and theoretical precision was observed at sampling rates of 20 MSPS, 40 MSPS, and 80 MSPS, as shown in Fig. 3. Subsequently, Gaussian pulses with the $\sigma$ of 4.3 ns and 1.2 ns, corresponding to FWHM of 10.127 ns and 2.826 ns, were also simulated under different SNRs and sampling rates, as shown in Table I. The simulation results show excellent agreement with the theoretical formula and this method achieves high time measurement precision that is proportional to the bandwidth of the pulse under test and the square root of the SNR. It should be noted that in practical applications, an appropriate sampling rate can be selected based on the bandwidth of the pulse to be measured and must satisfy the Nyquist sampling theorem.

As presented earlier in the theoretical analysis, the DFPF-based time measurement method is primarily determined by the similarity, bandwidth, and SNR of the digitized pulse waveform. In practical experimental applications, due to the influence of complex environments, the pulse waveform may not be identical, leading to slight variations in waveform similarity, shape, and some degree of SNR changes. Therefore, to better align with practical applications, baseline fluctuations, minor pulse shape variations, and different pulse amplitudes were introduced for further simulation experiments.

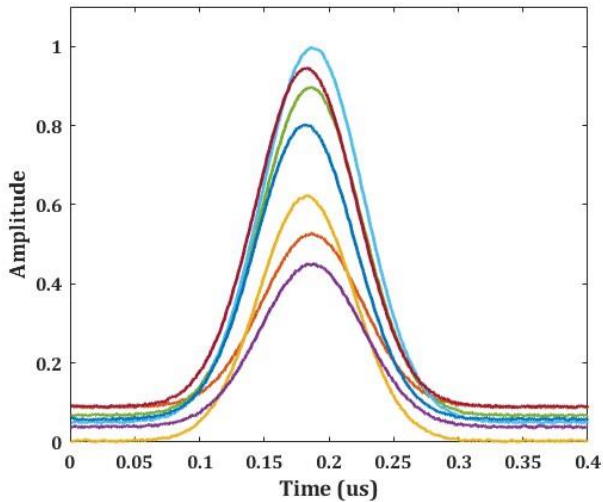

Fig. 4. A set of typical waveforms used for simulation, with baseline fluctuations, minor pulse shape variations, and different pulse amplitudes.

TABLE II
DFPF-BASED TIME MEASUREMENT PRECISION UNDER SPECIFIC SIMULATION CONDITIONS

| Pulse Width (ns) | SNR@A=1 (dB) | Random Amplitude Range | Precision (ps rms) | | | |
|---|---|---|---|---|---|---|
| | | | Tr0 | Tr1 | Tr2 | Tr3 |
| 117.750 | 64 dB | (0.20, 1.00) | 95 | 96 | 99 | 104 |
| | 50 dB | (0.20, 1.00) | 205 | 210 | 213 | 219 |
| | 40 dB | (0.20, 1.00) | 592 | 605 | 617 | 635 |
| | 30 dB | (0.20, 1.00) | 1780 | 1838 | 1982 | 2018 |
| 10.127 | 64 dB | (0.20, 1.00) | 41 | 56 | 60 | 74 |
| | 50 dB | (0.20, 1.00) | 84 | 92 | 113 | 152 |
| | 40 dB | (0.20, 1.00) | 235 | 247 | 262 | 284 |
| | 30 dB | (0.20, 1.00) | 899 | 913 | 993 | 1041 |
| 2.826 | 64 dB | (0.20, 1.00) | 5.7 | 7.8 | 10 | 12 |
| | 50 dB | (0.20, 1.00) | 14 | 17 | 19 | 20 |
| | 40 dB | (0.20, 1.00) | 45 | 47 | 48 | 51 |
| | 30 dB | (0.20, 1.00) | 135 | 146 | 151 | 158 |

The pulse widths in this simulation are 117.750 ns, 10.127 ns, and 2.826 ns, with corresponding sampling rates of 40 MSPS, 100 MSPS, and 500 MSPS, respectively. SNR@A represents the specific SNR of 64 dB, 50 dB, 40 dB, and 30 dB when the normalized amplitude is 1.00 as the reference. During the simulation, the pulse amplitudes are randomly generated between 0.20 and 1. Tr0, Tr1, Tr2, and Tr3 correspond to random rising edge fluctuations within the specified ranges of 100% to 100%, 100% to 105%, 100% to 110%, and 100% to 115%, respectively.

As shown in Fig. 4, a typical dataset from this simulation was presented, consisting of a Gaussian pulse with a width of approximately 117.750 ns. A normalized amplitude corresponding to specific SNR values of 64 dB, 50 dB, 40 dB, and 30 dB, respectively, with the normalized amplitude of 1.00 as the reference, was selected as the simulation data. During the simulation, the pulse amplitudes were randomly generated between 0.20 and 1. Subsequently, random low-frequency baseline drift ranging from 0% to 10% of normalized amplitude (1.00), along with random rising edge fluctuations within the specified ranges of 100 % to 105 %, 100 % to 110 %, and 100 % to 115 %, were introduced. Then, under the given constraints, two pulses with a fixed time interval were randomly selected from the aforementioned dataset to evaluate the time measurement precision using the DFPF-based method. Additionally, under similar conditions, simulations were also conducted by modifying the original pulse width to 10.127 ns and 2.826 ns.

The time measurement precision was derived from the statistical analysis of 50,000 events and the results were presented in Table II. The results show that minor baseline fluctuations have little effect on the SNR and similarity of the pulse waveform, and thus, have negligible impact on measurement precision. And minor pulse shape variations influence similarity, thereby affecting measurement precision slightly. However, changes in amplitude have a significant effect, primarily due to variations in SNR, as described in (20). As also shown in (20), the pulse width (FWHM in this paper) of different Gaussian pulses, representing different bandwidths, also significantly affects the time measurement precision.

## IV. EXPERIMENTS AND RESULTS

To validate the feasibility of the proposed method, a digital frequency-domain phase-fitting time measurement module (DFPF-TMM) is developed. Subsequently, an experimental platform is constructed, consisting primarily of the DFPF-TMM, a signal source, and a detector to assess its timing performance.

### A. Introduction to DFPF-TMM

The DFPF-TMM employs an ADC (ADS5263) for sampling nuclear pulse signals, while an FPGA (XC7K325TFFG900-2) serves as the digital processing unit, enabling time measurement based on the DFPF method. Fig. 5 presents the block diagram of the DFPF implementation on the FPGA, which primarily comprises the built-in fast Fourier transform (FFT) intellectual property (IP) core, the CORDIC IP core, and a custom-designed phase fitting module. First, the FFT IP core transforms the signal under test into the frequency domain. Next, the cross-correlation function is computed by multiplying the conjugation of one signal with another. Subsequently, the amplitude $A(k)$ and phase $P(k)$ of the cross-correlation function are extracted using the CORDIC IP core. In the coincidence module, a thresholding operation is applied to identify valid phase $P(k_n)$ data, which is then stored in the first-in, first-out (FIFO) buffer. Finally, linear regression is performed within the phase fitting module on the extracted phase data, with the resulting slope used to estimate the TOF information. The entire computation is implemented with fixed-point arithmetic to enhance processing efficiency and hardware compatibility.

Table III presents the resource utilization of the current version of the DFPF module, developed by our team, for a two-channel implementation on a Xilinx FPGA. This implementation supports a scalable 16-channel time measurement while utilizing 80% of the FPGA's available resources. The current version of the DFPF module operates at a 40 MHz clock. The delay introduced by the FPGA's FFT IP is 4.350 µs, while the CORDIC module incurs a delay of 0.325 µs. Additionally, the divider in the least-squares fitting module introduces a delay of 3.050µs. The total processing time is 20.675 µs for 512 samples.

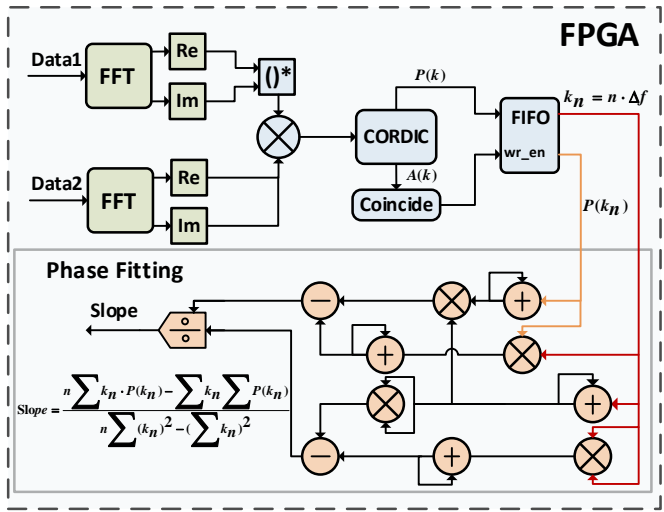

Fig. 5. Schematic diagram of the time measurement process implemented on an FPGA using the DFPF-based method.

TABLE III
FPGA RESOURCE UTILIZATION OF TWO-CHANNEL DFPF METHOD IMPLEMENTED ON A XC7K325TFFG900-2

| FPGA resources | Occupied | Available | Utilization |
|---|---|---|---|
| Occupied Slices | 5762 | 50950 | 11% |
| LUT | 11586 | 203800 | 6% |
| Slice Flip-flops | 20888 | 407600 | 5% |
| BRAM | 14.5 | 445 | 7% |
| DSPs | 52 | 840 | 6% |

### B. Performance of the Method

A testing platform was established in the laboratory using a Tektronix AFG31000 arbitrary waveform generator. Gaussian pulses were generated to simulate nuclear pulses and were fed into the DFPF-TMM to evaluate the performance of the proposed method.

The initial test conditions were as follows: the Gaussian pulse width was set to 118 ns, the TOF between pulses was 10 ns, the sampling rate was 40 MSPS, and 512 data points were input into the DFPF module for time extraction. The maximum achievable SNR for the input signals on this platform was 64 dB. The typical statistical results under this condition are shown in Fig. 6(a). The average measured time interval was 10.07 ns, with a STD of 44.6 ps RMS. Further analysis was conducted by introducing controlled noise into the test pulses. The timing precision of the proposed method was evaluated across varying input signal SNRs, ranging from 30 dB to 60 dB, and compared with the theoretical precision, as shown in Fig. 6 (b). When the input signal SNR drops below 40 dB, the method achieves a timing precision of approximately 600 ps. As the SNR further decreases to 30 dB, the precision degrades to 1.75 ns. As shown in Fig. 6(b), the theoretical precision is slightly better than the measured results, primarily due to the implementation of the DFPF algorithm on the FPGA using truncated fixed-point arithmetic to optimize resource utilization. The timing precision across a dynamic time interval range from 0 ns to 1 µs is presented in Fig. 7. The results indicate that under a 64 dB SNR and a Gaussian pulse width of 118 ns, the proposed method maintains a standard deviation between 40 ps and 50 ps RMS throughout the entire interval range, demonstrating the reliability and stability of the DFPF-TMM implementation.

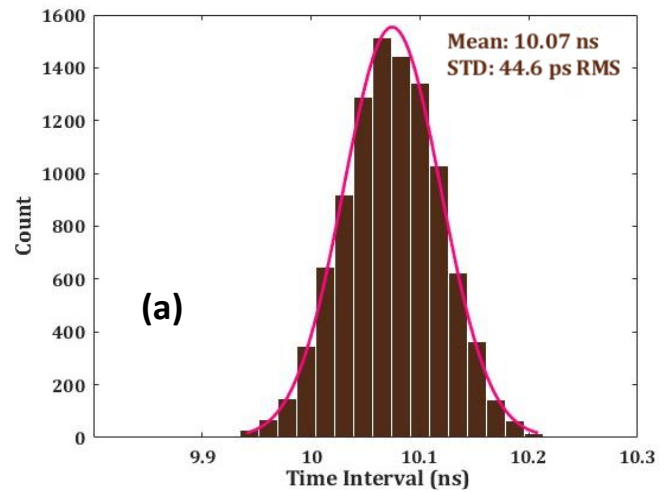

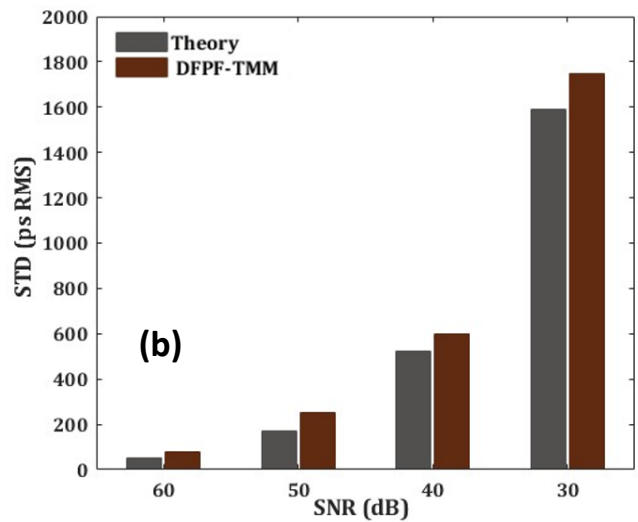

Fig. 6. (a) The typical histogram of the time measurement results under an SNR of 64 dB with a Gaussian pulse width of 118 ns. (b) The STD comparison between the theoretical values and the DFPF-TMM measurements under SNR conditions ranging from 30 dB to 60 dB, with a Gaussian pulse width of 118 ns.

Furthermore, Gaussian pulses with pulse widths of 10 ns and 3 ns, generated by the Tektronix AFG31000, were also tested, as presented in Table IV. The test results for Gaussian pulses with widths of 10 ns and 3 ns were first acquired by using an oscilloscope (RTM3002) and then imported into a computer for DFPF processing. The precision of the test results is slightly lower than that of the simulation due to the limited bit resolution of the oscilloscope. The results demonstrate that the proposed method is competitive with some of the best-reported values in literature. Table IV provides a comparison of experimental results, where the time precision reported by other authors, originally expressed as FWHM, has been converted to STD for consistency. Under similar conditions, all methods were evaluated using signal generators as the test signal source. The DFPF-based time measurement outperforms DCFD-based methods, including DCFD with time-domain interpolation or waveform fitting. This advantage primarily stems from the inherent averaging effect in the frequency-domain cross-correlation and phase-fitting process, which effectively minimizes measurement errors.

TABLE IV
COMPARISON OF DIFFERENT DIGITAL TIMING METHODS

| Method | Interpolation | Features | Pulse Width (ns) | fs (MSPS) | Precision (ps rms) | Implement On board | Ref |
|---|---|---|---|---|---|---|---|
| DCFD | FA | Non-linear fitting | ~10 | 100 | ~37 | Difficult | [20] |
| DCFD | WAA | Weighted average | ~10 | 100 | ~108 | Moderate | [20] |
| DCFD | FA | Non-linear fitting | ~10 | 250 | ~21 | Difficult | [20] |
| DCFD | WAA | Weighted average | ~10 | 250 | ~30 | Moderate | [20] |
| DCFD | Linear | Linear interpolation | ~3 | 500 | ~127 | Easy | [13] |
| DCFD | Cubic Spline | Cubic interpolation | ~3 | 500 | ~16 | Moderate | [13] |
| DCFD | Sinc | Sinc interpolation | ~3 | 500 | ~3.0 | Moderate | [13] |
| DFPF | / | Frequency-domain Linear fitting | ~10 | 100 | ~18 | Easy | This work |
| DFPF | / | Frequency-domain Linear fitting | ~10 | 250 | ~14 | Easy | This work |
| DFPF | / | Frequency-domain Linear fitting | ~3 | 500 | ~2.9 | Easy | This work |

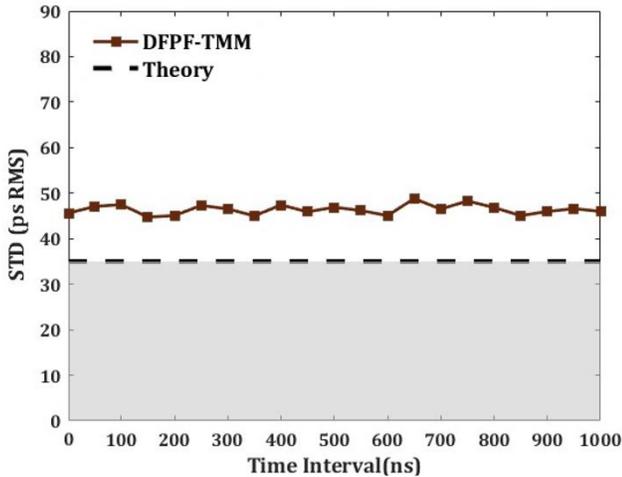

Fig. 7. Comparison of the measured STD of DFPF-TMM with the theoretical STD under the same SNR of 64 dB and pulse width of 118 ns, with time intervals ranging from 0 ns to 1 μs. Across this dynamic range, the STD measured by DFPF-TMM remains between 40 ps and 50 ps RMS, while the theoretical STD is 35 ps RMS

Furthermore, the time resolution of the DCFD-based method is inherently limited by the ADC sampling interval, although interpolation can improve precision to some extent. In contrast, by leveraging the DFPF approach, our method theoretically achieves an infinitely small time least significant bit (LSB). While this frequency-domain linear fitting method is slightly more complex than its time-domain linear fitting, the fitting process in the frequency domain involves an FFT and a single least-squares fit, making it more practical to implement than higher-order interpolation or advanced waveform fitting techniques. Moreover, it achieves high time precision and an excellent time LSB under Nyquist sampling conditions, offering a significant advantage in terms of both time resolution and computational efficiency.

*C. Cosmic Ray Test*

The block diagram of the cosmic ray test setup with the detector is shown in Fig. 8. The detector channel is based on a scintillator detector with a silicon photomultiplier (SiPM), comprising four key components: plastic scintillators, optical fibers, SiPMs, and charge sensitive amplifiers (CSAs) for channel readout [26]. In this study, the scintillator bars were manufactured by the GAONENGKEDI Company[27] using polyethylene extrusion technology [28]. The SiPM used is the S13360 series of multi-pixel photon counters (MPPCs) with a pixel size of 50 μm, produced by Hamamatsu. Additionally, the optical fibers were sourced from Kuraray [29].

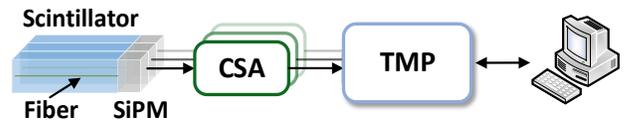

Fig. 8. The block diagram of the cosmic ray test. The scintillator, optical fiber, and SiPM form the scintillator-based detector.

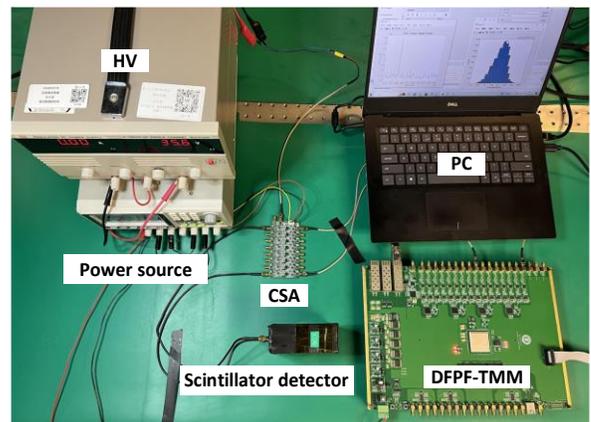

Fig. 9. The test platform for cosmic rays, including detectors, preamplifiers, electronics prototype, power supply, and display device.

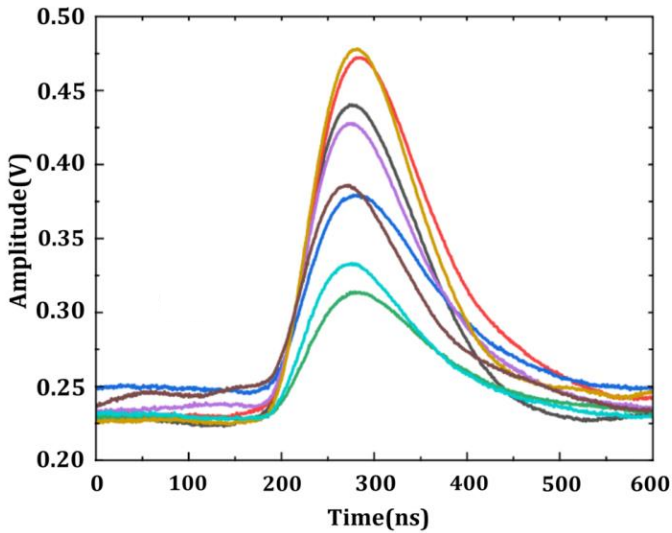

Fig.10. The oscilloscope-captured waveforms of cosmic rays. The oscilloscope is triggered under the condition that both channel 1 and channel 8 signals exceed a threshold of 270 mV simultaneously.

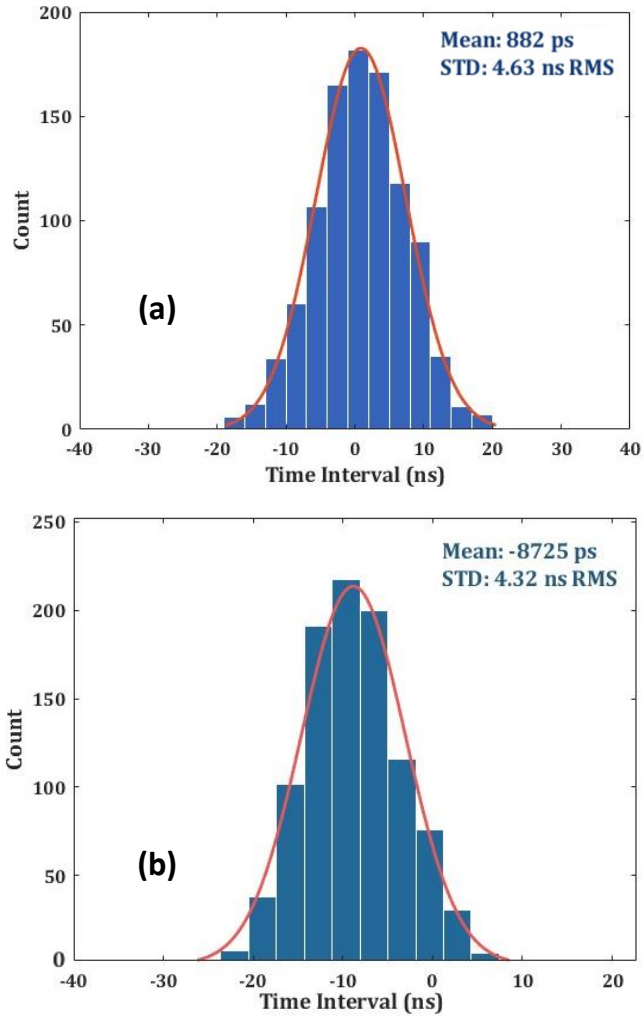

Fig. 11. (a). The histogram of the time measurement results in cosmic ray testing using the method based on DFPF. (b). The histogram of the time precision of the front-end detector.

The cosmic ray test platform is shown in Fig. 9. When the operating voltage of the detectors was set to the optimal value of 36 V, the cosmic ray signals, after amplification through the CSAs, were observed as shown in Fig. 10. Given that cosmic rays propagate at nearly the speed of light, the signals respond almost simultaneously, and the time intervals between signals from different channels can be approximated as 0 ns.

During the continuous testing exceeding 24 hours, more than 1000 valid cosmic ray events were collected. The histogram of the time measurement results from the cosmic ray test using this method is shown in Fig. 11(a), yielding a STD of 4.63 ns RMS. As illustrated in Fig. 11(b), the time measurement precision contribution of the detector was evaluated using a commercial instrument, DT5742, which resulted in 4.32 ns RMS. Considering the time precision contribution of the DT5742 instrument is 50 ps [30], this value is negligible. The SNR of the output cosmic ray pulses was approximately 30 dB. The results demonstrate that, under these conditions, the proposed method prototype contributes 1.7 ns RMS to the overall time precision, aligning well with the theoretical results presented in Fig. 3 and Table II in the previous section.

## V. CONCLUSION

This paper presents a high-precision time measurement method, which effectively utilizes the time information within digitized nuclear pulses by applying frequency-domain cross-correlation and phase-fitting techniques, implemented on an FPGA. Experimental results demonstrate that under an SNR of 64 dB, the proposed method achieves a time precision of 50 ps RMS for a Gaussian pulse with a width of approximately 118 ns at a 40 MSPS sampling rate. When the pulse width is reduced to approximately 3 ns at a 500 MSPS sampling rate, the precision improves to 2.9 ps RMS. The measured time precision is proportional to the signal bandwidth, as shown in (20), meaning it improves with a steeper rising edge.

The proposed method achieves high-precision time measurement under the condition that the output signals from the front-end detector and preamplifier circuit exhibit highly similar waveforms and have a bandwidth less than half of the ADC sampling rate, as required by the Nyquist sampling theorem. This method has the potential to serve as a valuable reference for both circuit board implementations and ASIC designs in time measurement applications. Furthermore, this method is not only suitable for nuclear pulse detection but also applicable to various other advanced time measurement scenarios.

ACKNOWLEDGMENT

The authors would like to thank Xiaolong Wang from Fudan University for his valuable experimental support. They also extend their gratitude to Qi Shen and Zhongtao Shen from the University of Science and Technology of China for their insightful discussions.